\documentclass[floatfix,aip,jcp]{revtex4-1}
\usepackage{multirow}
\usepackage{afterpage}
\usepackage{longtable}
\usepackage{xcolor}
\usepackage{array}
\usepackage{amsmath}
\usepackage{physics}
\usepackage{simpler-wick}
\usepackage{pdfcomment}
\usepackage{tikz}
\usepackage{mathtools}
\usepackage[utf8]{inputenc}
\usepackage[T1]{fontenc}
\usepackage{ifthen}
\usepackage[version=4]{mhchem}
 
\usepackage[top=2.0cm,left=1.5cm,right=1.0cm,bottom=2.25cm]{geometry}
\newcounter{noorthcounter} 
\newcounter{tdcounter} 
\newcounter{mocounter} 
\newcounter{tdmcscfcounter} 

\providecommand{\MO}{abc}
\renewcommand{\MO}
  {\addtocounter{mocounter}{1}\ifthenelse{\equal{\themocounter}{1}}{molecular orbital (MO)}{MO}}
 \providecommand{\MOs}{abc}
\renewcommand{\MOs}
  {\addtocounter{mocounter}{1}\ifthenelse{\equal{\themocounter}{1}}{molecular orbitals (MOs)}{MOs}}
\newcounter{aocounter} 
\providecommand{\AO}{abc}
\renewcommand{\AO}
  {\addtocounter{aocounter}{1}\ifthenelse{\equal{\theaocounter}{1}}{atomic orbital (AO)}{AO}}
 \providecommand{\AOs}{abc}
\renewcommand{\AOs}
  {\addtocounter{aocounter}{1}\ifthenelse{\equal{\theaocounter}{1}}{atomic orbitals (AOs)}{AOs}}
\newcounter{nocounter} 
\providecommand{\NO}{abc}
\renewcommand{\NO}
  {\addtocounter{nocounter}{1}\ifthenelse{\equal{\thenocounter}{1}}{natural orbital (\textit{no})}{\textit{no}}}
 \providecommand{\NOs}{abc}
\renewcommand{\NOs}
  {\addtocounter{nocounter}{1}\ifthenelse{\equal{\thenocounter}{1}}{natural orbitals (\textit{no}s)}{\textit{no}s}}
\newcounter{dodccounter} 
\providecommand{\DODC}{abc}
\renewcommand{\DODC}
  {\addtocounter{dodccounter}{1}\ifthenelse{\equal{\thedodccounter}{1}}{different-orbitals for different-configurations (DODC)}{DODC}}
\newcounter{vbcounter} 
\providecommand{\VB}{abc}
\renewcommand{\VB}
  {\addtocounter{vbcounter}{1}\ifthenelse{\equal{\thevbcounter}{1}}{valence bond (VB)}{VB}}
\newcounter{nocicounter}
\newcounter{nocisdcounter}
\providecommand{\NOCI}{abc}
\renewcommand{\NOCI}
  {\addtocounter{nocicounter}{1}\ifthenelse{\equal{\thenocicounter}{1}}{nonorthogonal configuration interaction (NOCI)}{NOCI}}
    \providecommand{\NOCISD}{abc}
\renewcommand{\NOCISD}
  {\addtocounter{nocicounter}{1}{\addtocounter{nocisdcounter}{1}\ifthenelse{\equal{\thenocicounter}{1}}{nonorthogonal configuration interaction singles and doubles (NOCISD)}{\ifthenelse{\equal{\thenocisdcounter}{1}}{NOCI singles and doubles (NOCISD)}{NOCISD}}}}
\newcounter{scfcounter} 
\providecommand{\SCF}{abc}
\renewcommand{\SCF}
  {\addtocounter{scfcounter}{1}\ifthenelse{\equal{\thescfcounter}{1}}{self-consistent field (SCF)}{SCF}}
\newcounter{mptwocounter} 
\providecommand{\MPTwo}{abc}
\renewcommand{\MPTwo}
  {\addtocounter{mptwocounter}{1}\ifthenelse{\equal{\themptwocounter}{1}}{M{\o}ller-Plesset second-order perturbation theory (MP2)}{MP2}}
\newcounter{mcscfcounter} 
\newcounter{nomcscfcounter} 
\newcounter{ssnomcscfcounter} 
\newcounter{sanomcscfcounter} 
\providecommand{\MCSCF}{abc}
\renewcommand{\MCSCF}
  {\addtocounter{mcscfcounter}{1}\ifthenelse{\equal{\themcscfcounter}{1}}{multiconfigurational self-consistent field (MCSCF)}{MCSCF}}
\providecommand{\TDNOMCSCF}{abc}
\renewcommand{\TDNOMCSCF}
  {\addtocounter{tdmcscfcounter}{1}\addtocounter{mcscfcounter}{1}\addtocounter{tdcounter}{1}\addtocounter{nomcscfcounter}{1}\addtocounter{noorthcounter}{1}\ifthenelse{\equal{\thetdcounter}{1}}{\ifthenelse{\equal{\themcscfcounter}{1}}{\ifthenelse{\equal{\thenoorthcounter}{1}}{real-time time-dependent nonorthogonal multiconfigurational self-consistent field (RT-TD NOMCSCF)}
      {real-time time-dependent NO multiconfigurational self-consistent field (RT-TD NOMCSCF)}}{\ifthenelse{\equal{\thenoorthcounter}{1}}{real-time time-dependent nonorthogonal MCSCF (RT-TD NOMCSCF)}{real-time time-dependent NOMCSCF (RT-TD NOMCSCF)}}}{\ifthenelse{\equal{\themcscfcounter}{1}}{\ifthenelse{\equal{\thenoorthcounter}{1}}{RT-TD nonorthogonal multiconfigurational self-consistent field (RT-TD NOMCSCF)}{RT-TD-NO multiconfigurational self-consistent field (RT-TD NOMCSCF)}}{\ifthenelse{\equal{\thenoorthcounter}{1}}{RT-TD nonorthogonal MCSCF (RT-TD NOMCSCF)}{RT-TD NOMCSCF}}}}

\providecommand{\TDMCSCF}{abc}
\renewcommand{\TDMCSCF}
  {\addtocounter{tdmcscfcounter}{1}\addtocounter{mcscfcounter}{1}\addtocounter{tdcounter}{1}\ifthenelse{\equal{\thetdcounter}{1}}{\ifthenelse{\equal{\themcscfcounter}{1}}{real-time time-dependent multiconfigurational self-consistent field (RT-TD MCSCF)}{real-time time-dependent MCSCF (RT-TD MCSCF)}}{ \ifthenelse{\equal{\themcscfcounter}{1}}{RT-TD multiconfigurational self-consistent field (RT-TD MCSCF)}{RT-TD MCSCF}}}

\providecommand{\NOMCSCF}{abc}
\renewcommand{\NOMCSCF}
  {\addtocounter{mcscfcounter}{1}\addtocounter{nomcscfcounter}{1}\addtocounter{noorthcounter}{1}\ifthenelse{\equal{\themcscfcounter}{1}}{nonorthogonal multiconfigurational self-consistent field (NOMCSCF)}{\ifthenelse{\equal{\thenomcscfcounter}{1}}{nonorthogonal MCSCF (NOMCSCF)}{NOMCSCF}}}
    \providecommand{\SSNOMCSCF}{abc}
\renewcommand{\SSNOMCSCF}
  {\addtocounter{mcscfcounter}{1}\addtocounter{nomcscfcounter}{1}\addtocounter{ssnomcscfcounter}{1}\addtocounter{noorthcounter}{1}\ifthenelse{\equal{\themcscfcounter}{1}}{state-specific nonorthogonal multiconfigurational self-consistent field (SS-NOMCSCF)}{\ifthenelse{\equal{\thenomcscfcounter}{1}}{state-specific nonorthogonal MCSCF (SS-NOMCSCF)}{\ifthenelse{\equal{\thessnomcscfcounter}{1}}{state-specifc NOMCSCF (SS-NOMCSCF)}{SS-NOMCSCF}}}}
    \providecommand{\SANOMCSCF}{abc}
\renewcommand{\SANOMCSCF}
  {\addtocounter{mcscfcounter}{1}\addtocounter{nomcscfcounter}{1}\addtocounter{sanomcscfcounter}{1}\addtocounter{noorthcounter}{1}\ifthenelse{\equal{\themcscfcounter}{1}}{state-average nonorthogonal multiconfigurational self-consistent field (SA-NOMCSCF)}{\ifthenelse{\equal{\thenomcscfcounter}{1}}{state-average nonorthogonal MCSCF (SA-NOMCSCF)}{\ifthenelse{\equal{\thesanomcscfcounter}{1}}{state-average NOMCSCF (SA-NOMCSCF)}{SA-NOMCSCF}}}}
\newcounter{casmaincounter} 
\newcounter{casscfcounter} 
\newcounter{sscasscfcounter} 
\newcounter{sacasscfcounter} 
\providecommand{\CAS}{abc}
\renewcommand{\CAS}
  {\addtocounter{casmaincounter}{1}\ifthenelse{\equal{\thecasmaincounter}{1}}{complete active space (CAS)}{CAS}}
  \providecommand{\CASSCF}{abc}
\renewcommand{\CASSCF}
  {\addtocounter{casmaincounter}{1}\addtocounter{casscfcounter}{1}\ifthenelse{\equal{\thecasmaincounter}{1}}{complete active space self-consistent field (CASSCF)}{\ifthenelse{\equal{\thecasscfcounter}{1}}{CAS self-consistent field (CASSCF)}{CASSCF}}}
    \providecommand{\SSCASSCF}{abc}
\renewcommand{\SSCASSCF}
  {\addtocounter{casmaincounter}{1}\addtocounter{casscfcounter}{1}\addtocounter{sscasscfcounter}{1}\ifthenelse{\equal{\thecasmaincounter}{1}}{state-specific complete active space self-consistent field (SS-CASSCF)}{\ifthenelse{\equal{\thecasscfcounter}{1}}{state-specific CAS self-consistent field (SS-CASSCF)}
{\ifthenelse{\equal{\thesscasscfcounter}{1}}{state-specific CASSCF (SS-CASSCF)}{SS-CASSCF}}}}
    \providecommand{\SACASSCF}{abc}
\renewcommand{\SACASSCF}
  {\addtocounter{casmaincounter}{1}\addtocounter{casscfcounter}{1}\addtocounter{sacasscfcounter}{1}\ifthenelse{\equal{\thecasmaincounter}{1}}{state-average complete active space self-consistent field (SA-CASSCF)}{\ifthenelse{\equal{\thecasscfcounter}{1}}{state-average CAS self-consistent field (SA-CASSCF)}
{\ifthenelse{\equal{\thesacasscfcounter}{1}}{state-average CASSCF (SA-CASSCF)}{SA-CASSCF}}}}
\newcounter{cicounter} 
\providecommand{\CI}{abc}
\renewcommand{\CI}
  {\addtocounter{cicounter}{1}\ifthenelse{\equal{\thecicounter}{1}}{configuration interaction (CI)}{CI}}
\newcounter{tdhfcounter} 
\providecommand{\TDHF}{abc}
\renewcommand{\TDHF}
  {\addtocounter{tdhfcounter}{1}\addtocounter{tdcounter}{1}\ifthenelse{\equal{\thetdcounter}{1}}{real-time time-dependent Hartree-Fock (RT-TDHF)} \ifthenelse{\equal{\thetdhfcounter}{1}}{RT-TD Hartree-Fock (RT-TDHF)}{RT-TDHF}} 
\newcounter{tdcicounter} 
\providecommand{\TDCI}{abc}
\renewcommand{\TDCI}
  {\addtocounter{tdcicounter}{1}\addtocounter{tdcounter}{1}\ifthenelse{\equal{\thetdcounter}{1}}{real-time time-dependent configuration interaction (RT-TDCI)}{\ifthenelse{\equal{\thetdcicounter}{1}}{RT-TD configuration interaction (RT-TDCI)}{RT-TDCI}}}
\newcounter{icmrcicounter} 
\providecommand{\icMRCI}{abc}
\renewcommand{\icMRCI}
  {\addtocounter{icmrcicounter}{1}\ifthenelse{\equal{\theicmrcicounter}{1}}{internally contracted multireference configuration interaction (ic-MRCI)}{ic-MRCI}}
\newcounter{onepdmcounter} 
\providecommand{\onePDM}{abc}
\renewcommand{\onePDM}
  {\addtocounter{onepdmcounter}{1}\ifthenelse{\equal{\theonepdmcounter}{1}}{one-particle density matrix (1PDM)}{1PDM}}
\newcounter{ticounter} 
\providecommand{\TI}{abc}
\renewcommand{\TI}
  {\addtocounter{ticounter}{1}\ifthenelse{\equal{\theticounter}{1}}{time-independent (TI)}{TI}}

\begin{document}
\title{Resolving Peak Shifting Effects in Real-Time Time-Dependent Orbital Propagation Methods Through Nonorthogonal Expansions}
\author{Matheus M. F. de Moraes}
\author{Lee M. Thompson}
\email{lee.thompson.1@louisville.edu}
\affiliation{Department of Chemistry, University of Louisville, 2320 South Brook Street, Louisville, KY 40292, USA}

\begin{abstract}
Self-consistent field real-time electronic structure methods are a powerful approach for modeling ultrafast processes but suffer from systematic errors in predicted transition energies, commonly known as peak shifting, that obscure direct comparison with experiment. Although this effect has been explored in single reference methods, a general understanding that extends to multiconfigurational methods remains unresolved. Here, we develop a real-time nonorthogonal multiconfigurational self-consistent field (RT-TD NOMCSCF) formalism in which independently propagated orbital sets generate a compact, nonorthogonal wavefunction. The approach establishes a unified framework for nonlinear real-time electronic structure theories, which we use to reveal the common origin of peak shifting across single and multireference limits. Using this framework, we demonstrate that peak shifting does not arise exclusively from the effect of state averaging over the internal space basis, but also from the accessible external configurational space. By systematically expanding the internal space with independently propagated orbital sets, RT-TD NOMCSCF suppresses this averaging, recovers correct transition energies, and distinguishes intrinsic peak shifting from numerical peak drifting caused by propagation errors. Thus, these developments establish nonlinear real-time electronic structure methods that enable more accurate simulations of ultrafast processes.
\end{abstract}

\maketitle

\section{Introduction}

Out-of-equilibrium ultrafast electron dynamics plays a central role in the exploration of electronic structure and photon-mediated processes.
Advances in ultrafast laser technology have enabled the development of powerful time-resolved (TR) experimental techniques that reveal previously inaccessible features of electronic and molecular structure. These include transient absorption (TA) spectroscopy,\cite{sekikawaPCCP23_25_8497,attarSci17_356_54} photoelectron spectroscopy,\cite{schuurmanPCCP22_24_20012}
long-range Rydberg-dressed interactions\cite{hollerithPRL22_128_113602}
 and high harmonic generation (HHG).\cite{Nisoli2017hr}
 The development of an accurate and affordable theoretical framework for modeling these processes faces many challenges.
 Initially, simulating time-dependent laser pulses requires a real-time time-propagation of the wavefunction, which drastically increases the computational cost in comparison to stationary state calculations.
A notable example is gas-phase X-ray pump-probe spectroscopy, where a core-to-valence or core-to-Rydberg excitation leads to a substantial relaxation of the wavefunction.
As a result, the transient states are not well modeled by the ground-state orbitals, and instead require orbitals optimized for either the relaxed core-exited state or the cation species.\cite{Oosterbaan2018io,Oosterbaan2019fa}
Therefore, despite many real-time time-dependent (RT-TD) methodologies reported over the years, only a few are sufficiently flexible to model electronic states far beyond the initial/ground state perturbative limit, as discussed in the following paragraphs.
Thus, the overall goal of this work is to develop and analyze a method that is able to resolve the challenge of describing real-time processes driven far from the initial system.

RT-TD methods can be divided into two classes -- those that just propagate the wavefunction through time-dependent state populations and those that also propagate the orbitals.
The former set consists of methods that propagate the population of the \TI{} states.\cite{schriberJCP19_151_171102}
Computationally, this approach is the most cost efficient as the bottleneck is the calculation, diagonalization and storage of the Hamiltonian and dipole matrices.
However, if the time propagation leads to large orbital relaxation, the initial state basis needs to be able to describe this relaxation. For orthogonal \CI{} based methods, accounting for orbital relaxation is typically achieved by using higher-order substituted configurations in the expansion, while in \CAS{} based methods a larger active space is required. Density Matrix Renormalization Group (DMRG) is another method in this family that can be used to significantly increase the size of the configuration expansion by performing the time propagation in a matrix-product-state representation.\cite{Wahyutama2024} 
Non-linear methods, such as RT-TD coupled-cluster methods, have a higher computational cost for the generation of the \TI{} matrices, require propagation of bra and ket wavefunctions due to their non-variational nature, and suffer from numerical instability when the ground state weight in the superposition tends to zero.\cite{pedersenJCP19_150_144106,ofstadWCMS23_13_e1666}
Alternatively, methods based on a nonorthogonal determinant expansion, where the orbitals of each configuration are optimized independently, are a viable strategy to reduce the expansion required to describe multiple states with similar accuracy.\cite{Dong.2024}
However, all methods using a fixed \TI{} state expansion in the propagation are limited, both by the initial expansion choice which requires the user to foresee the states involved and can lead to user bias in the calculation, as well as the fact that many properties such as peak position and transition probability can be obtained prior to the time-dependent propagation.

Using orbital propagation techniques, the user bias in the state basis expansion can be minimized. After the initial wavefunction is defined, both orbitals and state populations (if multiconfigurational) can modify the configuration space based on the time-dependent Schr{ö}dinger equation.
The class of methods based on \TDMCSCF{} provides a fully variational and more flexible framework in which both the configuration coefficients and the underlying single-particle orbitals are time-dependent and optimized simultaneously according to the time-dependent variational principle.\cite{katoCPL04_392_533,satoPRA13_88_023402,miyagiPRA13_87_062511}
This approach allows the orbital basis itself to adapt dynamically to external perturbations such as laser-driven evolution, enabling a more compact wavefunction to sample a larger manifold of the Hilbert space, although at a higher computational cost. 
However, a consequence of orbital time propagation is that the final properties are dependent on both initial guess and applied electric field.
For example, \TDHF{}, which is the single reference limit of \TDMCSCF{}, can provide accurate results in the limit of weak perturbations.
When the field cannot be considered as a weak perturbation of one of the stationary states, the density and the associated properties behave as a weighted average of multiple states.
Two representative examples are two-level Rabi oscillations\cite{habenichtJCP14_141_184112} and the transition energies obtained from the Fourier transform of the dipole, in an effect known as peak-shifting.\cite{provorseJCTC15_11_4791}
In both cases, the system propagation behaves as a two-electron process, with resonant energies proportional to the double excited density and its relative energy.\cite{isbornJCP08_129_204107}
This effect can be mitigated by increasing the size of the  reference space, which allows for changes in the population of multiple internal states without changes in the orbitals.
However, peak-shifting is still observed in \TDMCSCF{} methods, where it is referred to as time-dependent state averaging.\cite{padmanabanCPL08_463_263} 
The state average is a generalization of the \TDHF{} density averaging, where the orbitals experience different forces from distinct configurations. 
To reduce this averaging effect larger internal spaces are required, as changes in state populations can be used to accommodate the effect of multiple configurations on the orbital derivatives.

An alternative way to minimize time-dependent state averaging is to simultaneously and independently propagate multiple sets of orbitals.
This approach substantially increases the flexibility for defining the internal space and, as a result, reduces the time-dependent total density averaging.
In general, this approach leads to configurations in the determinant basis that are nonorthogonal, even when an orthogonal expansion is used initially.
Nonorthogonal wavefunction methods have been proposed as a compact way to model near-degenerate configurations with substantially distinct optimal sets of orbitals.\cite{Kempfer-Robertson.2022olv}
Therefore, we propose the development of a \TDNOMCSCF{} approach that propagates distinct orbital sets and \CI{} coefficients together to provide a more robust internal space and minimize averaging artifacts observed in \TDMCSCF{} and \TDHF{}. The remainder of this article outlines our pilot implementation and explores the behavior of the method.

\section{Theory}
\label{sec:theory}
There are multiple ways to derive a \TDNOMCSCF{} methodology.
In this section, we follow an internally-contracted approach, where a single transformation in a common basis is applied to the superposition as a whole.
An alternative configuration-specific formulation using a Dirac-Frenkel time-dependent variational principle can be found in Appendix A.

Considering a normalized initial superposition written in terms of a linear combination of nonorthogonal Slater determinants, 
\begin{align}
    |\Psi(t)\rangle = \sum_I |I(t)\rangle C_I(t) \label{eq:initial_state},
\end{align}
where each configuration $|I(t)\rangle$ is a product of orthogonal orbitals contained in independent configuration-specific basis sets, $^I\phi_p(t) \in {^IB}(t)$. As all terms are dependent on time, we drop the explicit time variable for the sake of clarity in the remainder of the derivation.
At any time step, this superposition derivative with respect to time can be represented as all single excitations over an arbitrary common orbital basis, $\phi_{\tilde u} \in \tilde B$, and based on the time-dependent Schr{ö}dinger equation we have, 
\begin{align}
    i\frac{\partial |\Psi\rangle}{\partial t} =  i\sum_{\tilde u, \tilde v} |\Psi_{\tilde u}^{\tilde v}\rangle \eta_{\tilde u}^{\tilde v}={\hat H}|\Psi\rangle \label{eq:time_dependent_Seq},
\end{align}
where  $\eta_{\tilde u}^{\tilde v}$ is the orbital gradient between $\tilde u$ and $\tilde v$.
For nonorthogonal expansions, an excitation acting on one of the determinants will, in general, produce a configuration with non-zero projection on both the original internal space (ie. $[\{|I\rangle\}]$) and external space (ie. the full space after removal of the internal space).\cite{moraesJCP26_164_014107} 
Therefore, it is necessary to remove redundancies that result from the action of the parameters $\eta_{\tilde u}^{\tilde v}$ caused by the ill-defined excitation mapping.

The parameters that only change the phase of the time-dependent superposition can be obtained by projecting eq.~\refeq{eq:time_dependent_Seq} over the superposition,
\begin{align}
      i\sum_{\tilde u, \tilde v} |\Psi\rangle\langle\Psi|\Psi_{\tilde u}^{\tilde v}\rangle \eta_{\tilde u}^{\tilde v}=|\Psi\rangle\langle\Psi|{\hat H}|\Psi\rangle =|\Psi\rangle E \label{eq:proj_state},
\end{align}
where $E$ is the energy of the superposition. 
Subsequently, the phase component of $\eta_{\tilde u}^{\tilde v}$ can be obtained by projecting eq.\ \ref{eq:proj_state} over a single excited wave function and multiplying both sides by the pseudo-inverse of the projected single-single excitation overlap matrix,
\begin{align}
      i{{}^\circ \eta_{\tilde r}^{\tilde s}}=i\sum_{\tilde u, \tilde v,\tilde p, \tilde q} ({^\circ S_{\tilde r\tilde s,\tilde p\tilde q}})^+({^\circ S_{\tilde p\tilde q,\tilde v\tilde u}})\eta_{\tilde u}^{\tilde v}=\sum_{\tilde p, \tilde q} ({^\circ S_{\tilde r\tilde s,\tilde p\tilde q}})^+\langle\Psi_{\tilde p}^{\tilde q}|\Psi\rangle E,\label{eq:circ_eta}
\end{align}
where ${^\circ S_{\tilde p\tilde q,\tilde v\tilde u}}=\langle\Psi_{\tilde p}^{\tilde q}|\Psi\rangle\langle\Psi|\Psi_{\tilde u}^{\tilde v}\rangle$ is the single-single excitation overlap matrix, $+$ indicates the Moore–Penrose pseudo-inverse, and ${{}^\circ \eta_{\tilde r}^{\tilde s}}$ is the non-redundant phase-component of the propagation contained in the excitation ${\tilde r}\to{\tilde s}$.

Similarly, to obtain the internal space component of the superposition, eq.~\eqref{eq:time_dependent_Seq} is projected over all configurations in eq.~\eqref{eq:initial_state} using the projector $\hat P=\sum_{I, J}|I\rangle\langle I|J\rangle^{+}\langle J|$.
The projector requires the inverse of the overlap matrix due to the nonorthogonality among configurations, and the pseudo-inverse is used to allow propagation near singularities in the internal space. 
Therefore, the non-redundant internal component is given by
\begin{align}
      i{{}^\parallel \eta_{\tilde r}^{\tilde s}}=i\sum_{\tilde u, \tilde v,\tilde p, \tilde q} ({^\parallel S_{\tilde r\tilde s,\tilde p\tilde q}})^+({^\parallel S_{\tilde p\tilde q,\tilde v\tilde u}})\eta_{\tilde u}^{\tilde v}=\sum_{\tilde p, \tilde q} ({^\parallel S_{\tilde r\tilde s,\tilde p\tilde q}})^+\langle\Psi_{\tilde p}^{\tilde q}|\bigg(\hat P-|\Psi\rangle\langle\Psi|\bigg)\hat H|\Psi\rangle,\label{eq:par_eta}
\end{align}
where ${^\parallel S_{\tilde p\tilde q,\tilde v\tilde u}}=\langle\Psi_{\tilde p}^{\tilde q}|\bigg(\hat P-|\Psi\rangle\langle\Psi|\bigg)|\Psi_{\tilde u}^{\tilde v}\rangle$.
Finally, the external space component is obtained in the same manner as,
\begin{align}
      i{{}^\perp \eta_{\tilde r}^{\tilde s}}=i\sum_{\tilde u, \tilde v,\tilde p, \tilde q} ({^\perp S_{\tilde r\tilde s,\tilde p\tilde q}})^+({^\perp S_{\tilde p\tilde q,\tilde v\tilde u}})\eta_{\tilde u}^{\tilde v}=\sum_{\tilde p, \tilde q} ({^\perp S_{\tilde r\tilde s,\tilde p\tilde q}})^+\langle\Psi_{\tilde p}^{\tilde q}|\bigg(1-\hat P\bigg)\hat H|\Psi\rangle,\label{eq:perp_eta}
\end{align}
with ${^\perp S_{\tilde p\tilde q,\tilde v\tilde u}}=\langle\Psi_{\tilde p}^{\tilde q}|\bigg(1-\hat P\bigg)|\Psi_{\tilde u}^{\tilde v}\rangle$.
Therefore, combining eqs.\ \eqref{eq:circ_eta}, \eqref{eq:par_eta} and \eqref{eq:perp_eta}, it is possible to define the non-redundant space-separated time derivative as,
\begin{align}
    \frac{\partial |\Psi\rangle}{\partial t} =\sum_{\tilde u, \tilde v} \bigg(1-\hat P\bigg)|\Psi_{\tilde u}^{\tilde v}\rangle {^\perp\eta_{\tilde u}^{\tilde v}} + \sum_{\tilde u, \tilde v} \bigg(\hat P-|\Psi\rangle\langle\Psi|\bigg)|\Psi_{\tilde u}^{\tilde v}\rangle {^\parallel\eta_{\tilde u}^{\tilde v}}+\sum_{\tilde u, \tilde v} |\Psi\rangle\langle\Psi|\Psi_{\tilde u}^{\tilde v}\rangle {^\circ \eta_{\tilde u}^{\tilde v}}\label{eq:state_dt}.
\end{align}

These equations, in general, result in $ {\eta_{\tilde u}^{\tilde v}} \neq  -{\eta_{\tilde v}^{\tilde u}}^*$ and consequently break the orbital-orthogonality of the common basis $\tilde B$.
By enforcing orthogonality in the orbital basis (${^IB}$) of each Slater determinant $|I\rangle $, the change in the metric matrix of the common orbital basis is translated into changes in the internal space metric matrix elements, $^{IJ}S=\langle I\vert J\rangle$.
As a result, an initially orthogonal determinant expansion can evolve over time into a nonorthogonal expansion.

The coupled propagation of all independent basis sets is obtained by mapping the common basis gradient to each ${}^{I}B$ and to each \CI{} coefficient.
This mapping is obtained by substituting the time-derivative of eq.\ \eqref{eq:initial_state} into the left hand side of eq.\ \eqref{eq:state_dt}:
\begin{align}
    \sum_I \sum_{i,a}|I_i^a\rangle {^I\eta_i^a}C_I+ \sum_I |I\rangle \dot C_I=\sum_{\tilde u, \tilde v} \bigg(1-\hat P\bigg)|\Psi_{\tilde u}^{\tilde v}\rangle {^\perp\eta_{\tilde u}^{\tilde v}} + \sum_{\tilde u, \tilde v} \bigg(\hat P-|\Psi\rangle\langle\Psi|\bigg)|\Psi_{\tilde u}^{\tilde v}\rangle {^\parallel\eta_{\tilde u}^{\tilde v}}+\sum_{\tilde u, \tilde v} |\Psi\rangle\langle\Psi|\Psi_{\tilde u}^{\tilde v}\rangle {^\circ \eta_{\tilde u}^{\tilde v}}\label{eq:config_to_state_dt},
\end{align}
where ${^I\eta_i^a}$ defines rotation between orbitals $^I\phi_i$ and $^I\phi_a$, such that  ${^I\eta_i^a}= -{^I\eta_a^i}^*$, and $\dot C_I$ is the time derivative of the state coefficient associated with $|I\rangle$.
Projecting eq.~\eqref{eq:config_to_state_dt} over an external single-excited configuration, the orbital derivative can be uncoupled from the coefficient propagation,
\begin{align}
    \sum_I \sum_{{\tilde u},{\tilde v}}\langle J_{\tilde p}^{\tilde q}|\bigg(1-\hat P\bigg)|I_{\tilde u}^{\tilde v}\rangle {^I\eta_{\tilde u}^{\tilde v}}C_I =& \sum_{\tilde u, \tilde v}\langle J_{\tilde p}^{\tilde q}|\bigg(1-\hat P\bigg) |\Psi_{\tilde u}^{\tilde v}\rangle {^\perp\eta_{\tilde u}^{\tilde v}},\label{eq:com_eta}
 \end{align}   
 here we use the fact that the single excitation space is independent from the basis choice,\cite{moraesJCP26_164_014107} and the first term of the left hand side of eq.~\eqref{eq:config_to_state_dt} can be rewritten as $\sum_I \sum_{{\tilde u},{\tilde v}}|I_{\tilde u}^{\tilde v}\rangle {^I\eta_{\tilde u}^{\tilde v}}C_I$.
 Therefore, the common orbital gradients over $|I\rangle$ are
 \begin{align}
    {^I\eta_{\tilde u}^{\tilde v}} =& \frac{C_I^*}{|C_I|^2}\sum_{L,K}\sum_{\tilde p, \tilde q}\sum_{\tilde r, \tilde s} {^{\perp IL} S_{\tilde u\tilde v,\tilde q\tilde p}^+}{^{\perp LK} S_{\tilde p\tilde q,\tilde r\tilde s}}C_K {^\perp\eta_{\tilde s}^{\tilde r}}\label{eq:dCIdt}
\end{align}
or zero if $|C_I|^2 \le \epsilon$, where $ \epsilon$ is a threshold sufficiently close to zero (in this work $ \epsilon=10^{-7}$). This threshold arises from the case where the configuration has no weight in the superposition, so that its orbitals have no effect on the total energy causing their propagation to be undetermined. By setting the undetermined orbital parameter propagation to zero, the determinant remains relevant only through propagation of its \CI{} coefficient, so that a determinant with zero weight in the initial superposition can obtain nonzero weight during the time propagation.
The molecular orbital rotations of each determinant set are then computed using the basis transformation matrices  (${^IT_{\tilde up}}$) and forcing anti-Hermiticity,
\begin{align}
   {^I\eta_{i}^{a}} = \sum_{{\tilde u},{\tilde v}} {^IT_{\tilde va}^*}{^I\eta_{\tilde u}^{\tilde v}}{^IT_{\tilde ui}}\text{ and }  {^I\eta_{a}^{i}} \coloneq -{^I\eta_{i}^{a*}},
\end{align}
where $i$ and $a$ are the occupied and virtual orbital indices in $|I\rangle$, respectively.
Each transformation generates an arbitrary phase in each set of orbitals, but the relative phases among sets of orbitals, which have an impact on the coupling elements, are dictated by the common basis transformation.
The effect of the relative phase is explored in further detail in Sec.~\ref{sec:results_space} and Appendix~B.

Once the MO specific propagation is known, the coefficient terms are computed by projecting eq.~\eqref{eq:config_to_state_dt} over the complement of the internal space, 
\begin{align}
    \sum_I  \langle J|\bigg(\hat P-|\Psi\rangle\langle\Psi|\bigg)|I\rangle {\dot C_I}=& \sum_I\sum_{\tilde u, \tilde v}\langle J|\bigg(\hat P-|\Psi\rangle\langle\Psi|\bigg) |I_{\tilde u}^{\tilde v}\rangle\bigg( {^\parallel\eta_{\tilde u}^{\tilde v}} - {^I\eta_{\tilde u}^{\tilde v}}\bigg)C_I\label{eq:coef_dt}\\
    {\dot C_K}=&  \sum_{IJ}  \sum_{\tilde u, \tilde v}\Bigg[\langle K|\bigg(\hat P-|\Psi\rangle\langle\Psi|\bigg)|J\rangle\Bigg]^+\langle J|\bigg(\hat P-|\Psi\rangle\langle\Psi|\bigg) |I_{\tilde u}^{\tilde v}\rangle\bigg( {^\parallel\eta_{\tilde u}^{\tilde v}} - {^I\eta_{\tilde u}^{\tilde v}}\bigg)C_I.\nonumber
\end{align}
In general, this transformation is non-unitary, as ${^\parallel\eta_{\tilde u}^{\tilde v}} -{^I\eta_{\tilde u}^{\tilde v}}$ accounts for changes in the metric matrix caused by the independent MO propagation. 
The orbitals and coefficient of the following time step are then obtained via a unitary transformation and linear propagation, respectively.
The fourth-order Runge-Kutta (RK4) method was used for the time propagation to reduce the numerical error.

At any time step, the time-dependent stationary states, $|\Phi_A\rangle=\sum_I |I\rangle D_{IA}$, can be computed by solving the generalized eigenvalue problem of the associated time-dependent Hamiltonian and overlap matrices:
\begin{align}
   \mathbf{HD} = \mathbf{DSE},
\end{align}
for ${}^{IJ}H = \langle I|\hat H|J\rangle$ and ${}^{IJ}S = \langle I|J\rangle$.
Lastly, the superposition population of each time-dependent state is defined by the projection,
\begin{align}
  p_A(t)  = \langle\Phi_A|\Psi\rangle =  \sum_{IJ} D^*_{IA}{}^{IJ}SC_{J}.
\end{align}
Additionally, the nonorthogonal matrix element machinery can be used to the compute autocorrelation function, defined as\cite{padmanabanCPL08_463_263} 
\begin{align}
  \langle\Psi(0)|\Psi\rangle =  \sum_{IJ} C^*_{I}(0)\langle I(0)|J\rangle C_{J}.
\end{align}

The results reported use a non-truncated set of natural-excitations, that is, all orbitals in the common basis are considered active and all excitations are computed.~\cite{moraesJCP26_164_014107}
Such a choice increases the overall number of redundant excitations, but removes the dependence on the orbital representation and orbital subspace definitions and does not affect the scaling of the step which is the computational bottleneck, as discussed in Appendix~C.

\section{Results}
\subsection{Effect of Initial Superposition and Internal Space}
\label{sec:results_space}

\begin{figure}
    \centering
    \includegraphics{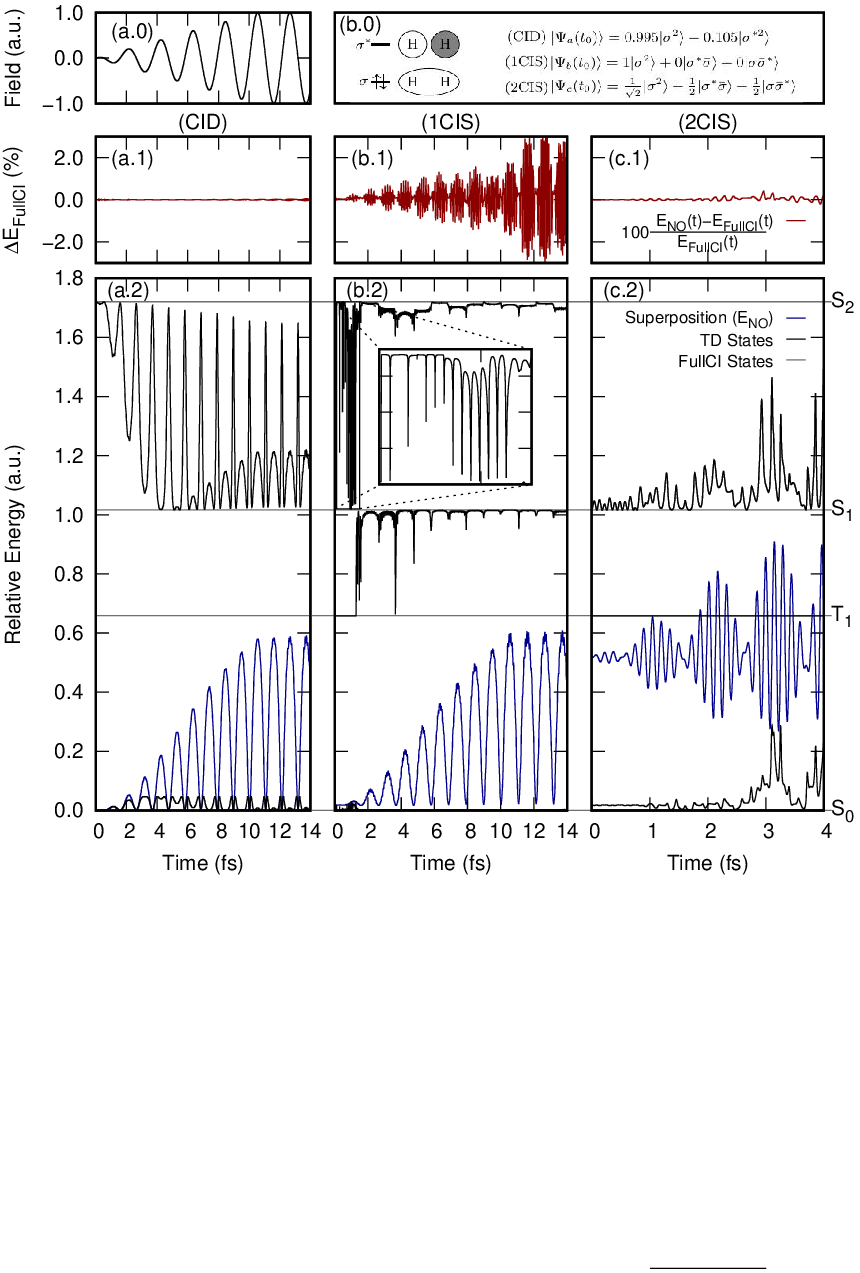}
    \caption{H$_2$ energy over time (in fs) under oscillatory field with initial ramp function, frequency of 0.072 a.u. and max amplitude of 1.0 a.u. (as depicted in a.0).
    The initial orbital space, composed of two restricted orbitals (STO-3G basis set  at the equilibrium distance), is shown in panel (b.0) together with
    three initial orthogonal expansions, $|\Psi_x(t_0)\rangle$ ($x$=a,b,c), written over their complete internal spaces, here denoted by CID, 1CIS and 2CIS.
    The column label indicates the initial condition and internal space used for the results below it.
    In Panels (x.1) the percentage error of the proposed RT-TD NOMCSCF superposition energy (E$_\text{NO}$) with respect to the full CI propagation (E$_\text{FullCI}$) of $|\Psi_x(t_0)\rangle$.
    In Panels (x.2) RT-TD NOMCSCF time-dependent relative energies with respect to the full CI ground state for the superposition (E$_\text{NO}$ in blue) and time-dependent states (black).
    Horizontal gray lines marks the energy of the full CI states (labels on the right).
     Inset in panel (b.2) shows zoom into region between 1.0 and  1.7 a.u. energy for the first 1.5 fs.}
    \label{fig:H2_Comp_dipole}
\end{figure}

As for any multiconfigurational method, the accuracy of the proposed approach is strongly dependent on the initial sets of orbitals and configurations. The small Hilbert space generated by \ce{H2} using a minimal basis, comprising three singlet states and one triplet state, makes it a good model system to explore the properties of \TDNOMCSCF{}. As the full CI wavefunction has four dimensions (with three independent variables), a \TDNOMCSCF{} expansion with two determinants should be sufficient to exactly reproduce any superposition state.

Figure~\ref{fig:H2_Comp_dipole} shows the time-dependent energy of three distinct initial wavefunctions (as summarized in panel~b.0) under a very strong oscillatory electric field with a five optical cycle ramp, as shown in panel~a.0. 
The simulation in the first column (labeled CID) uses the minimal expansion required for the full CI ground state (i.e.\ RHF reference plus its double-excited configuration, $|\Psi_a(t_0)\rangle$). 
Columns 1CIS and 2CIS are both expansions comprising of single excitations over the RHF reference along with the RHF reference itself, where the initial superposition in the former (center column) is the \TI{} ground state, $|\Psi_b(t_0)\rangle$, while the latter (right column) starts from an equal mix of ground and first excited singlet, $|\Psi_c(t_0)\rangle$. 
In all cases, the initially orthogonal determinant sets rapidly evolve into nonorthogonal configurations.
Although the spin symmetry of the closed-shell configurations in the CID expansion is maintained, the magnitude and phase of the coupling oscillates in response to the field.
As a consequence, the internal space is a time-dependent two-dimensional sub-space of the three-dimensional singlet state-space, spanned by $S_{0}$, $S_{1}$ and $S_{2}$. 
This effect can be observed in Fig.~\ref{fig:H2_Comp_dipole}~(a.2), where the lowest time-dependent state energy is constrained between the full CI ground state and the energy of a slightly polarized restricted single determinant, while the second time-dependent state has the full CI $S_1$ state energy as its floor and the full CI $S_2$ state energy as its ceiling. 
This superposition propagation follows full CI as shown by the near zero energy error in panel (a.1). The small oscillatory error is caused by the higher sensitivity to the cumulative numerical errors of the non-linear orbital propagation and the non-unitary nature of the linear \CI{} propagation, as described in the discussion of eq.~\eqref{eq:coef_dt}.

During the 1CIS and 2CIS trajectories, the open-shell determinants results in  $ {\eta_{\tilde 1\alpha}^{\tilde 2\alpha}}={\eta_{\tilde 1\beta}^{\tilde 2\beta}}\neq{\eta_{\tilde 2\alpha}^{\tilde 1\alpha}}={\eta_{\tilde 2\beta}^{\tilde 1\beta}}$.
In an orthogonal method these orbital gradients would be average-weighted to generate a unitary transformation over the common orbital basis.
As discussed during eq.~\refeq{eq:config_to_state_dt}, the proposed method allows for each determinant to propagate based on the superposition, but without average-weighting their orbital gradients.
The determinant-specific gradients are $ {^{\sigma^*\bar\sigma}\eta_{1\alpha}^{2\alpha}}={^{\sigma\bar\sigma^*}\eta_{1\beta}^{2\beta}}\neq {^{\sigma^*\bar\sigma}\eta_{1\beta}^{2\beta}}={^{\sigma\bar\sigma^*}\eta_{1\alpha}^{2\alpha}}$, i.e.\ the open-shell determinant orbital propagation is mirrored because of the initial spin-symmetry conditions.
This symmetry results from how the method conserves the spin-symmetry of the superposition, without requiring the spin-symmetry to be enforced using a constraint over either the orbitals or determinant basis.
The overlap between these two open-shell determinants is twice the unsigned occupied $\alpha$-$\beta$ spatial orbital overlap. Therefore, the observed spin-symmetry conservation is only obtained by permitting the superposition to evolve as a nonorthogonal expansion.

As a consequence of the \TDNOMCSCF{} spin-symmetry conservation, solving the time-dependent eigenvalue problem during the 1CIS and 2CIS trajectories should always yield the triplet state with constant energy along with two time-dependent singlet states that span the evolving superposition.
This effect can be observed in the first femtosecond of panel~(b.2) and panel~(c.2), where the triplet state has a constant energy while the singlet states change in energy.
During the initial attoseconds of the 1CIS trajectory, the coefficients of the open-shell configurations are zero (as shown in $|\Psi_b(t_0)\rangle$ panel b.0), causing eq.~\eqref{eq:dCIdt} to be set to zero as the wavefunction propagation is invariant to orbital changes in these determinants.
The field interaction slowly populates the $S_1$ state, such that once the coefficients of the open-shell determinants surpass $\epsilon$, the orbital time derivatives start to be computed.
This leads to an abrupt orbital and state energy change, but not a discontinuity, around 20 as (see inset in panel~(b.2)).
For sufficiently small time-steps the ground state smoothly propagates from the RHF solution to the full CI ground state, while the excited singlet state oscillates between $S_1$ and $S_2$. Despite the abrubt change in the internal space, the total energy and dipole remain continuous in this region.

Although, the propagation is physically sound, the sudden energy change upon the open-shell configuration weights reaching the threshold has a major impact on the numerical stability.
As discussed in Sec.~\ref{sec:theory}, the relative phase between orbital sets is propagated via eq.~\eqref{eq:dCIdt} and, in general, is associated with the electric field frequency (e.g. as observed in Fig.~\ref{fig:H2_Comp_dipole}~(a.0) and (a.2)).
In the trajectory shown in panel (b.2), the initial sudden change in orbitals generates an ultra-fast time-dependent relative phase (as shown in the inset), which causes steep derivatives in the state energies every couple of 0.1 fs and, consequently, fast growth in the cumulative error, as shown in panel~(b.1).
This error eventually breaks the time-dependent states total-spin symmetry and creates a spin-contaminated internal space.
After the break in symmetry, the relative phases among determinants change, which causes one of the states to vary between a singlet and a triplet in the same frequency as the electric field.
The large cumulative numerical error, combined with the total-spin symmetry breaking,  disturbs the propagation to the point where the wavefunction is a completely distinct superposition to that of the full CI simulation, which is the cause of the apparently large relative error shown in panel~(b.1).
It is relevant to note that, despite the spin-contaminated internal space and its effect over the propagation, the superposition total-spin remains a pure singlet during the entire propagation.

It is important to highlight that the behavior of the ground state in the 1CIS trajectory is caused by the initial space and superposition combination, rather than the initial internal space choice alone.
The 2CIS trajectory shown in Fig.~\ref{fig:H2_Comp_dipole}~(c.2) exemplifies this point, as the reference configurations are the same but instead the initial superposition is an equal weight of ground and first excited singlet state ($|\Psi_c(t_0)\rangle$ panel b.0).
During this propagation, the 2CIS total superposition energy (blue line) oscillates with a higher frequency than the 1CIS simulation.
The opposite behavior is observed for the state energies (black lines), where the singlet state energies for 2CIS change more slowly than in the 1CIS trajectory.
Figure~\ref{fig:H2_Comp_dipole}~(c.1) shows a much smaller error relative to full CI than panel (b.1). Additionally, the error increase occurs at time steps where the state energy derivative rapidly changes. These observations suggest that the cumulative error arises primarily from the numerical propagation failing to capture higher order derivatives of the time-dependent state energies, rather than the superposition energy.

In summary, we find that provided the \TDNOMCSCF{} equations contain sufficient numbers of parameters, the time propagation follows that of full CI propagation, demonstrating the correctness of the derivation and implementation. The combination of initial internal space and superposition weights has a large impact on the numerical robustness of the propagation. 

\subsection{Spectral Peak Shifting}

As discussed during the Introduction, the superposition obtained by field-mediated orbital rotations is known for averaging transition energies.
This effect has been studied using H$_2$ with a minimal basis in detail,\cite{provorseJCTC15_11_4791} such that it is a suitable system for testing the robustness of \TDNOMCSCF{} against peak shifting.
In this prior work, starting from the restricted DFT ground state, a short electric-field pulse was found to mediate the mixing between $\sigma$ and $\sigma^*$ restricted orbitals (initial orbitals depicted in Fig.~\ref{fig:H2_Comp_dipole} panel b.0), which generated a field-dependent $S_0$ and $S_2$ superposition state.
The Fourier transform of the \ce{H2} RT-TDDFT dipole after the pulse end showed a single averaged $S_0+S_2\to S_1$ transition in the spectrum, rather than the expected pair of transitions ($S_0\to S_1$ and $S_2\to S_1$).

\begin{figure}
    \centering
    \includegraphics{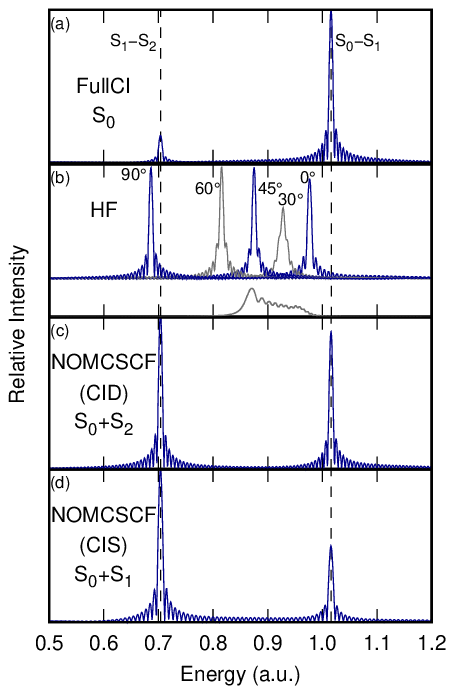}
    \caption{H$_2$ STO-3G transition spectra computed via dipole Fourier transformed real-time propagation. 
    Vertical dashed line indicates the energy of the time-independent full CI transitions label on the top and RT-TD full CI in Panel~(a) for a ground state initial superposition.
    Panel~(b) summarizes RT-TDHF peak-averaging effects. 
    On the top, peak-shifting shown by superimposing the results of orbital-rotated initial references (labels indicates rotation).
    On the bottom, peak-drifting shown by propagating the RHF ground state over a long time with large time-steps.
    Panels (c) and (d) contain RT-TD NOMCSCF results for CID and CIS expansion, respectively.
    The initial superposition for both (c) and (d) is an equal mix between ground and excited singlet states.
    }
    \label{fig:H2_shift}
\end{figure}

Fig.~\ref{fig:H2_shift}~(a) presents reference results for this system obtained using full CI.
The propagation used to generate these results begins with the population entirely in the ground state. 
An initial electric-field pulse partially populates the first excited state, making both allowed transitions accessible, i.e.\ $S_1\to S_2$ (0.7038~a.u.) and $S_0\to S_1$ (1.015~a.u.), so that the corresponding peaks are present in the spectrum.

Fig.~\ref{fig:H2_shift}~(b) depicts multiple \TDHF{} results and exemplify two effects (one methodological and the other numerical) that affect the peak position of non-linear RT-TD methods. 
In the top portion of panel~(b), we show the peak shifting effect which arises from the orbital and state-averaging over the spectra, also observed in RT-TDDFT calculations.
In these calculations, the $\sigma$ and $\sigma^{*}$ orbitals are rotated from 0$^\circ$ to 90$^\circ$ (as indicated in the figure), preserving spin symmetry, before the field was applied.
The orbital and state averaging generated by each rotation leads to transition energy averaging proportional to the rotation angle.
The bottom of panel~(b) shows peak drifting, which is a consequence of the numerical error inherent to orbital- and density-propagation based methods.
Over long trajectories, the cumulative error gradually mixes the internal and external spaces, resulting in a smooth displacement of the spectral peaks.
This error can first be spotted by a non-physical change in the total energy in absence of an external electric field.
In the calculation shown in the bottom panel of Fig.~2~(b), this numerical error has intentionally been induced by propagating the orbitals linearly using a large time step.
As a consequence, the initial aufbau density converts over time into an ionic state with the occupied restricted-orbital localized over one of the hydrogen atoms, similar to the 45$^\circ$ rotated reference.
The transition energy follows the changing superposition composition, resulting in the signal smearing over a band of energies. In contrast, numerical error when propagating only state populations without orbital propagation impacts only the signal intensity and not the positions.

Using the peak-shifting behavior observed for single-reference methods and the exact full CI results as a reference, we can investigate how the \TDNOMCSCF{} method affects the electron dynamics, and so impacts the observed peak shifting in the simulation.
Panel~(c) of Fig.~\ref{fig:H2_shift} shows the results obtained by propagating a wavefunction initially constructed from equal weights of S$_0$ and S$_2$ states, i.e. a minimal internal space, generated using the CID expansion.
During the propagation, the internal-space depends on time, which generates time-dependent states.
Therefore, the relative energy between S$_0(t)$ and S$_2(t)$ oscillates over time and is dependent on the initial pulse. 
Equal peak positions are obtained regardless of the initial internal space and superposition weights, as illustrated by tests propagating of the 2CIS superposition (shown in panel~(d)), $1/\sqrt{2}\vert\sigma^{2}\rangle+1/\sqrt{2}\vert (1/\sqrt{2}\sigma-1/\sqrt{2}\sigma^*)^{2}\rangle$ (not shown), and $1\vert{\sigma\bar{\sigma}^{*}}\rangle+0\vert{\bar{\sigma}\sigma^{*}}\rangle$ (not shown).
In all tested cases, the transition peaks remain in agreement with the full CI reference. 
This observation indicates that the peak-averaging effect, including both peak shifting and peak drifting, does not arise solely from state averaging in the propagation of the occupied orbitals over the states contained within the internal space. It also reflects an analogous averaging over the manifold of external states generated by single excitations from the reference states.

\begin{figure}
    \centering
    \includegraphics{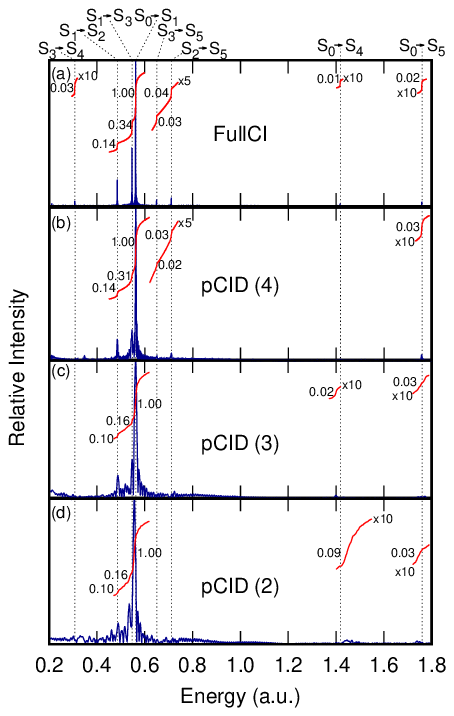}
    \caption{H$_2$ 6-31G transition spectra computed via dipole Fourier transformed real-time propagation. 
    Vertical dashed line indicates the energy of the time-independent full CI transitions label on the top.
    Panel (a) real-time RT-TD full CI results.
    Panels (b) to (d) contain the results obtained from real-time RT-TD NOMCSCF calculations with a selection  of initial expansions with includes the RHF ground state and: (b) all orthogonal-pair-double excitations (pCID); (c) its two lowest energy orthogonal-pair-double excitations; and (d) its first lowest energy orthogonal-pair-double excitation.
    Inset curves in red shows the cumulative-integral of the spectra, the relative area below each peak is represented by the values next to it, while any scaling is marked by the x-labels.
    }
    \label{fig:H2_6_31G}
\end{figure}

To explore the effect of the change in the accessible external space caused by the orbital-propagation on the energies of the spectral signals, the basis set was increased to 6-31G, while all other parameters were kept the same. 
Fig.~\ref{fig:H2_6_31G}~(a) shows the full CI reference for the 6-31G basis set.
Each of the relevant transitions are generated by the six low-lying roots out of the ten total singlet states, where S$_0$ ($|1\sigma_g^2\rangle$), S$_2$ (spin-adapted $|1\sigma_g^12\sigma_g^1\rangle$) and S$_3$ ($|1\sigma_u^2\rangle$) have $^1\Sigma_g^+$ symmetry, and  S$_1$ (spin-adapted  $|1\sigma_g^11\sigma_u^1\rangle$), S$_4$ (spin-adapted  $|1\sigma_g^12\sigma_u^1\rangle+|2\sigma_g^11\sigma_u^1\rangle$) and S$_5$ (spin-adapted $|1\sigma_g^12\sigma_u^1\rangle-|2\sigma_g^11\sigma_u^1\rangle$) have $^1\Sigma_u^+$ symmetry.
Inspired by the results shown in Fig.~\ref{fig:H2_shift}~(c), we explored the viability of a seniority-zero initial expansion (i.e. all possible determinants with zero unpaired electrons).
The largest internal space explored was composed of four determinants and started from the full seniority zero expansion (i.e. all pair-double excitations).
The results (panel~b) are in good agreement with full CI in both intensity and position of the signals.
Therefore, by considering only restricted closed-shell configurations in the internal space, both closed-to-open-shell and open-to-open-shell transitions can be accurately modeled.
The main disagreement is the S$_1\to$S$_3$ peak intensity, which results from peak broadening due to small numerical propagation error causing peak drifting.
The effect of peak drifting can be mitigated by comparing the relative peak area (shown with red curves), rather than the maximum intensity. 

Removing the highest energy $|2\sigma_u^2\rangle$ determinant from the full seniority-zero internal space does not significantly change the spectrum profile when the initial superposition is the S$_0$ state, as shown in Fig.~\ref{fig:H2_6_31G}~(c).
he main peaks $S_0\to S_1$, $S_1\to S_3$  and  $S_1\to S_2$ are all present at a similar energy to that obtained from the trajectories using the more complete expansions, although with less accurate intensities. Despite the reduction of the accessible external space resulting from the smaller reference expansion having a small impact on the peak position, changes over time in the number of linearly independent excitations have a much larger effect on the numerical stability of the propagation.
As a consequence, the overall error is tied to the threshold used for removing redundant external configurations in the pseudo-inverse calculation of ${}^{\parallel}S_{\tilde{r}\tilde{s},\tilde{p}\tilde{q}}$, ${}^{\perp}S_{\tilde{r}\tilde{s},\tilde{p}\tilde{q}}$ and ${}^{\perp IJ}S_{\tilde{r}\tilde{s},\tilde{p}\tilde{q}}$ in eqs.~\eqref{eq:par_eta}, \eqref{eq:perp_eta} and \eqref{eq:dCIdt}.
The effect of the pseudo-inverse threshold is accentuated by reductions in the size of the internal space, which explains why spectra computed with fewer determinants are generally noiser, i.e.\ there is less defined separation between relevant and redundant external space configurations.
As an example, an initial expansion composed of configurations $|1\sigma_g^2\rangle$ and $|1\sigma_u^2\rangle$ shows a strong peak shift depending on the psuedo-inverse threshold.
For a reasonable threshold (10$^{-8}$ in this case), as shown in Fig~\ref{fig:H2_6_31G}~(d), the peak shifting is less noticeable, while the broadening effect caused by peak drift is apparent.
Peak shifting is mainly present in the $S_0\to$S$_5$ transition, likely due to the high multireference character of the S$_5$ state, as shown by comparing the peak positions, where the \TDHF{} peak is at 1.633 a.u.\ ($\pm0.005$ a.u.), while two-, three- and four-determinant \TDNOMCSCF{} give 1.735 a.u.\ (+0.02/-0.002 a.u., i.e.\ peak drifting to lower energies), 1.770 a.u.\ (+0.005/-0.03 a.u., i.e.\ peak drifting to higher energies) and 1.758 a.u.\ ($\pm0.001$) respectively, compared to 1.7585 a.u.\ for full \CI{}.
These results indicate that the peak-averaging effect occurs in \TDNOMCSCF{} propagation but is rapidly reduced as the internal space increases, while the numerical stability of the method is not directly proportional to the internal expansion size.

\begin{figure}
    \centering
    \includegraphics{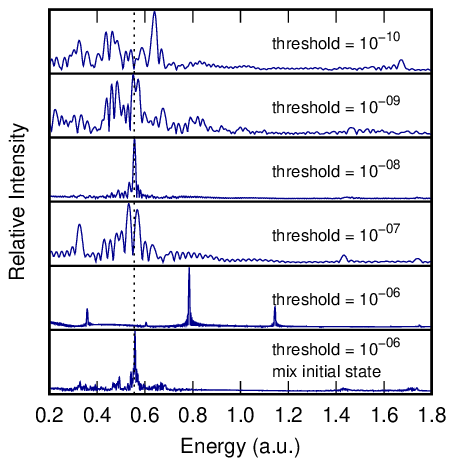}
    \caption{H$_2$ 6-31G transition spectra computed via dipole Fourier transformed from real-time RT-TD NOMCSCF calculations with a selection of initial expansions with includes the RHF ground state and its first lowest energy orthogonal-pair-double excitation. 
    Each panel shows the result obtained using a pseudo-inverse threshold (as labeled).
    All propagation starts from a pure ground state, with the exception of the panel labeled as "mix initial state", which is composed of 75\% ground state and 25\% first excited state.
    }
    \label{fig:pinv}
\end{figure}

The dependence of the time-propagation stability on the pseudo-inverse threshold is undesirable, but not uncommon in non-linear multi-reference internally contracted based approaches, such as internally-contracted multireference coupled cluster.\cite{evangelistaJCP11_134_114102}
On the other hand, the impact of such arbitrary threshold allows us to change the accessible external space without modifying the internal space in order to explore the effect of the external space size on peak shifting.
Fig.~\ref{fig:pinv} shows the dependence of the result in Fig.~\ref{fig:H2_6_31G}~(d) on the pseudo-inverse threshold.
Thresholds that are too small, such as $10^{-9}$ and $10^{-10}$ fail to remove linearly-dependent external configurations.
This effect can be easily observed, as it causes the total energy to keep varying after the electric field is turned off.
For thresholds that are too high, the propagation behavior is sound, but an increase in peaks shifting and cumulative error occurs. 
Using the most intense peak as example ($S_{0}\rightarrow S_{1}$), it shifts from 0.56 a.u.\ (threshold = 10$^{-8}$) to 0.53/0.57 a.u.\ (10$^{-7}$), and then to 0.79 a.u.\ (10$^{-6}$).
If the pseudo-inverse threshold is sufficiently high, only a small, but constant, external sub-space is accessible. 
By obtaining a constant external space dimension through the high pseudo-inverse threshold, it is possible to reduce the numerical error in the calculation even though the result shows incorrect physical behavior, as observed in the simulation using a 10$^{-6}$ pseudo-inverse threshold starting in the ground state.
The main issue in choosing an acceptable threshold for computing the pseudo-inverse is its dependence on the initial expansion. 
For instance, no peak shifting is observed using the same two initial determinants as discussed above with a threshold of 10$^{-6}$ if an initial superposition composed of 75\% ground state and 25\% first excited state is used (bottom panel of Fig.~\ref{fig:pinv}). 
\section{Conclusion}

In this work, we presented the equations and a pilot implementation of a flexible \TDNOMCSCF{} approach, in which the state populations are propagated in conjunction with configuration-specific orbitals, with the orbital set associated with each configuration propagated independently.
By projecting the total superposition orbital time-derivative into orbital-specific components, the occupied orbitals of each configuration can be propagated based on the interaction among all configurations, but without the orbital-averaging imposed by a common molecular-orbital basis set.
As a consequence, the basis of configurations naturally propagates into the nonorthogonal regime, allowing the wavefunction to be represented in a compact form.

Using \ce{H2} as a model system, we explored the origin of the peak-shifting problem that plagues single-reference RT-TD methods, as well as the related time-dependent state-averaging problem, which affects the spectral energy values of RT-TD multireference methods.
Our results indicate that this effect does not arise only from the deficiencies or limitations imposed in the construction of internal space, but also from the accessible external sub-space, which is defined by the excitations on top of the evolving superposition.
We also explored a more prevalent effect in non-linear propagation, the peak-drifting effect, where the cumulative error leads to small variations in the transition energies.
The main consequence of peak drifting is transition broadening, in contrast to peak shifting which alters the transition energy.
Although a more robust numerical integration method is the best solution for mitigating peak drifting, both the relative energies and intensities can be accurately determined using the peak integral, even where peak drifting occurs.

Finally, we demonstrated how the proposed \TDNOMCSCF{} method is able to accurately model transitions to, from and between open-shell dominant states using a small expansion of pair-excited configurations.
As a consequence, this method can sample the most relevant states in the external space based on the TD Schr{ö}dinger equation.
This is a large advantage in comparison to RT-TD methods that only propagate state populations, whose properties are entirely defined by the initial set of \TI{} states, and as a result, heavily dependent on the user's choice of configuration expansion.
Additionally, the results indicate that nonorthogonally-coupled seniority-zero expansions have the potential for enabling RT-TD simulations with reduced dependence on the internal space construction.

\section{Acknowledgments}
This work was supported by the U.S. Department of Energy, Office of Science, Basic Energy Sciences, in the Computational and Theoretical Program (Grant No. DE-SC0024507). 

\section{Data Availability}
The data that support the findings of this study are available from the corresponding author upon reasonable request.

\section{Appendix A: Dirac–Frenkel–McLachlan Formulation of the RT-TD NOMCSCF Equations}

Similar \TDNOMCSCF{} equations can be obtained without the common basis and using the Dirac-Frenkel-McLachlan TD variational principle, described by Miyagi and Madsen.\cite{miyagiPRA13_87_062511}
In this case the action functional is defined as:
\begin{align}
  S&[\{C_I\},\{^I\phi_i\},\{^I\epsilon_i^j\},\lambda]=\\
  &\int_0^T\Bigg[\langle\Psi|\Bigg(i\frac{\partial}{\partial t}-H\Bigg)|\Psi\rangle+\lambda\bigg(\langle\Psi|\Psi\rangle-1\bigg)+\sum_I\sum_{ij}{^I\epsilon_i^j}\bigg(\langle^I\phi_j|^I\phi_i\rangle-\delta_{ij}\bigg)\Bigg],\nonumber
\end{align}
where $\lambda$ is the Lagrange multiplier that ensures normalization of the wavefunction expansion coefficients, while ${^I\epsilon_i^j}$ ensures orthonormality of the orbital set associated with configuration $|I\rangle$.

Expanding the wavefunction using eq.~\eqref{eq:initial_state}, the stationary point of the functional is defined as:
\begin{align}
  \delta S&[\{C_I\},\{^I\phi_i\},\{^I\epsilon_i^j\},\lambda]=\\
  &\int_0^T\Bigg[\sum_{IJ}\Bigg(\delta C_I^*\langle I|+ C_I^*\langle\delta I|\Bigg)\Bigg(i|\dot J\rangle C_J+i|J\rangle \dot C_J-H|J\rangle C_J\Bigg)\nonumber\\
  &-\Bigg(i\dot C_I^*\langle I|+ iC_I^*\langle\dot I|+C_I^*\langle I|H\Bigg)\Bigg(|\delta J\rangle C_J+|J\rangle \delta C_J\Bigg)\nonumber\\
  &+\delta\lambda\bigg(\langle\Psi|\Psi\rangle-1\bigg)+\lambda\sum_{IJ}\bigg(\delta C_I^*\langle I|J\rangle C_J+ C_I^*\langle\delta I|J\rangle C_J+ C_I^*\langle I|\delta J\rangle C_J+ C_I^*\langle I|J\rangle \delta C_J\bigg)\nonumber\\
  &+\sum_I\sum_{ij}\delta{^I\epsilon_i^j}\bigg(\langle^I\phi_j|^I\phi_i\rangle-\delta_{ij}\bigg)+\sum_I\sum_{ij}{^I\epsilon_i^j}\bigg(\langle\delta^I\phi_j|^I\phi_i\rangle+\langle^I\phi_j|\delta^I\phi_i\rangle\bigg)\Bigg]=0.\nonumber
\end{align}
Applying the stationary condition for a variation in the coefficients ($\partial S/\partial C_I^*=0$) and expanding the time derivative using eq.~\eqref{eq:config_to_state_dt} we obtain
\begin{align}
  &\sum_{J}\Bigg(i\sum_{jb}\langle I|J_j^b\rangle C_J{^J\eta_{jb}}+i\langle I|J\rangle \dot C_J-\langle I|H|J\rangle C_J\Bigg)+\lambda\sum_{J}\langle I|J\rangle C_J=0.\label{apx:ci_der}
\end{align}
Following the same procedure for the variation of an occupied orbital in $|I\rangle$ and projecting into a virtual orbital (i.e. $\langle^I\phi_a|\partial S/\partial \langle^I\phi_i|=0$) 
\begin{align}
  &\sum_{J}\Bigg( i\sum_{jb}C_I^*\langle I_i^a|J_j^b\rangle C_J{^J\eta_{jb}}+iC_I^*\langle I_i^a|J\rangle \dot C_J-C_I^*\langle I_i^a|H|J\rangle C_J\Bigg)\label{apx:orb_der}\\
  &+\lambda\sum_{J}C_I^*\langle I_i^a|J\rangle C_J+\sum_{j}{^I\epsilon_j^i}\langle^I\phi_a|^I\phi_j\rangle=0,\nonumber
\end{align}
where the last term on the left-hand side is zero due to orbital orthogonality within the determinant $I$. As a result, the only remaining multiplier is $\lambda$ in both equations. 

Dividing the $\lambda$ dependent term of eq.~\eqref{apx:orb_der} into internal and external components using the identity operator ($1-\hat P +\hat P$), a term with a similar dependence on $\lambda$ as eq.~\eqref{apx:ci_der} is obtained,
\begin{align}
  \lambda\sum_{J}C_I^*\langle I_i^a|J\rangle C_J=\lambda\sum_{J}C_I^*\langle I_i^a|1-\hat P|J\rangle C_J +C_I^*\langle I_i^a|\hat P|J\rangle C_J =  \sum_{JKL}C_I^*\langle I_i^a|K\rangle^{KL}S^+\langle L|J\rangle C_J \lambda.\label{apx:lambda_term}
\end{align}
Substituting eq.~\eqref{apx:ci_der} into \eqref{apx:lambda_term} and then substituting the result into eq.~\eqref{apx:orb_der} gives us the orbital rotation terms
\begin{align}
  &\sum_{J}\Bigg[ i\sum_{jb}C_I^*\langle I_i^a|\bigg(1-\hat P\bigg)|J_j^b\rangle C_J{^J\eta_{jb}}+iC_I^*\langle I_i^a|\bigg(1-\hat P\bigg)|J\rangle \dot C_J-C_I^*\langle I_i^a|\bigg(1-\hat P\bigg)|H|J\rangle C_J\Bigg]=0\nonumber\\
  \Rightarrow & \sum_{J}i\sum_{jb}C_I^*\langle I_i^a|\bigg(1-\hat P\bigg)|J_j^b\rangle C_J{^J\eta_{jb}}= \sum_{J} C_I^*\langle I_i^a|\bigg(1-\hat P\bigg)|H|J\rangle C_J,
\end{align}
which has a similar form to equations~\eqref{eq:perp_eta} and \eqref{eq:com_eta}. Multiplying both sides by the pseudo-inverse of the left-hand side overlap term, the isolated orbital derivative has the same form as eq.~\eqref{eq:dCIdt}, also depending on the inverse of the coefficients norm.

The linear coefficient derivatives can be obtained by multiplying eq.~\eqref{apx:ci_der} by all  expansions orthogonal to the superposition of interest ($|\Psi\rangle$), that is,
\begin{align}
  &\sum_{J}\Bigg[i\sum_{jb}\langle I|\bigg(\hat P -|\Psi\rangle\langle\Psi|\bigg)|J_j^b\rangle C_J{^J\eta_{jb}}+i\langle I|\bigg(\hat P -|\Psi\rangle\langle\Psi|\bigg)|J\rangle \dot C_J-\langle I|\bigg(\hat P -|\Psi\rangle\langle\Psi|\bigg)|H|J\rangle C_J\Bigg] =0\nonumber\\
  \Rightarrow &\sum_{J}i\langle I|\bigg(\hat P -|\Psi\rangle\langle\Psi|\bigg)|J\rangle \dot C_J=\sum_{J}\Bigg[\langle I|\bigg(\hat P -|\Psi\rangle\langle\Psi|\bigg)|H|J\rangle C_J -i\sum_{jb}\langle I|\bigg(\hat P -|\Psi\rangle\langle\Psi|\bigg)|J_j^b\rangle C_J{^J\eta_{jb}}\Bigg].\label{apx:ci_result}
\end{align}
In this equation, the first term of the RHS accounts for the propagation itself, while the second term accounts for changes in the internal space metric matrix due to the orbital rotations, as in eq.~\eqref{eq:coef_dt}.

\section{Appendix B: phase-consistent biorthogonalization scheme}

During the \TDNOMCSCF{} propagation each orbital in each basis set representation has a constantly evolving arbitrary phase.
While this arbitrary phase has no effect in the diagonal terms of the Hamiltonian, the relative phase among sets or orbitals can have a huge impact in the coupling elements and as such must be correctly considered.
The standard approach to compute the nonorthogonal coupling elements starts with the biorthogonalization of the two sets of occupied molecular orbitals using single value decomposition (SVD).\cite{Mahler.2021,Dong.2024}
In this approach, the occupied-molecular-orbital overlap between configurations is transformed by,
\begin{align} 
     \sum_{\nu\tau} {^IC^\dagger_{i\nu}}g_{\nu\tau}{^JC_{\tau j}}= \sum_p {^{IJ}U_{ip}}{^{IJ}\Sigma_{p}}{^{IJ}V^\dagger_{pi}},
\end{align}
where $g_{\nu\tau}$ is the atomic orbital overlap between orbitals $\nu$ and $\tau$, $U$ and $V$ define the transformations such as ${^J\mathbf{\tilde C}}={^J\mathbf{C}}\mathbf{V}$ and ${^I\mathbf{\tilde C}}={^I\mathbf{C}}\mathbf{U}$, and ${^{IJ}\mathbf{\Sigma}}$ is a diagonal matrix containing the overlap between the transformed orbitals.

Although a useful approach, the SVD transformed orbitals have arbitrary relative sign, which can cause sign indeterminacy in real-orbitals\cite{Mahler.2021} and arbitrary phase in complex-orbitals.
Because the same phase is applied in both the configuration overlap and the Hamiltonian coupling, the arbitrary phase is not a problem if the determinant pair has non zero overlap.
On the other hand, pairs of non orthogonal configurations with zero overlap are phase dependent.
In the real case, the arbitrary phase is just a sign that can be easily corrected by multiplying the matrix element by the sign of the transformation matrix determinants, i.e. $\text{sgn}(\det(\mathbf{U})\det(\mathbf{V}))$.

In the complex case, to avoid the relative phase problem among complex-orbitals the biorthogonal transformations can be obtained by solving the following eigen problem,
\begin{equation}
    \sum_{\nu\tau} \sum_i {^IC^\dagger_{j\nu}}g_{\nu\tau}{^JC_{\tau i}}V_{iq}= \sum_i V_{ji}s_{iq},
\end{equation}
where $\mathbf{V}$ is the eigenvector matrix and $\mathbf{s}$ contains the biorthogonal orbital overlaps up to a phase.
If $\mathbf{{{}^I C^\dagger}g{{}^J C}}$ is normal (ie. it commutes with its complex conjugate), $\mathbf{V}$ is unitary and transforms both molecular-orbitals, ${^J\mathbf{\tilde C}}={^J\mathbf{C}}\mathbf{V}$  and ${^I\mathbf{\tilde C}}={^I\mathbf{C}}\mathbf{V}$.
However, this approach is dependent on the arbitrary orbital order, that is, the matrix can be non-normal, but made normal by an orbital order swapping.
This problem can be better exemplified in an orthogonal case, consider two electrons  in the spin-orbital basis $\{\phi_1,\phi_2,\phi_3,\phi_4\}$.
The occupied orbital overlap between  $|\phi_1\phi_2\phi_3\rangle$ and $|\sqrt{0.5}(\phi_1-\phi_3)\sqrt{0.5}(\phi_1+\phi_3)\phi_4\rangle$ is given by
\begin{align}
\mathbf{{{}^I C^\dagger}g{{}^J C}}=\mathbf{A}=&\begin{bmatrix}
\frac{1}{\sqrt{2}} & -\frac{1}{\sqrt{2}} &0\\
0 & 0  &0\\
\frac{1}{\sqrt{2}} &\frac{1}{\sqrt{2}}& 0.\label{apB:ex1}
\end{bmatrix}
\end{align}
Using SVD we obtain the expected two pairs of orbitals overlaps with norm one ($\{\phi_1,\phi_3\}$ and $\{\sqrt{0.5}(\phi_1-\phi_3)\sqrt{0.5}(\phi_1+\phi_3)\}$) and a pair with zero overlap ($\{\phi_2\}$ and $\{\phi_4\}$), 
\begin{align}
\begin{bmatrix}
\frac{1}{\sqrt{2}} & -\frac{1}{\sqrt{2}} &0\\
0 & 0  &0\\
\frac{1}{\sqrt{2}} &\frac{1}{\sqrt{2}}& 0
\end{bmatrix}=
\begin{bmatrix}
\frac{1}{\sqrt{2}} & -\frac{1}{\sqrt{2}} &0\\
0 & 0  &1\\
\frac{1}{\sqrt{2}} &\frac{1}{\sqrt{2}}& 0
\end{bmatrix}
\begin{bmatrix}
1 & 0 & 0\\
0 & 1 & 0\\
0 & 0 & 0
\end{bmatrix}\begin{bmatrix}
1 & 0 & 0\\
0 & 1 & 0\\
0 & 0 & 1
\end{bmatrix}.\nonumber
\end{align}
In this case, $\mathbf{A}$ is not normal ($\mathbf{A^\dagger A}\neq\mathbf{AA^\dagger} $) and can be at most decomposed as a Jordan matrix,
\begin{align}
\begin{bmatrix}
\frac{1}{\sqrt{2}} & -\frac{1}{\sqrt{2}} &0\\
0 & 0  &0\\
\frac{1}{\sqrt{2}} &\frac{1}{\sqrt{2}}& 0
\end{bmatrix}
\begin{bmatrix}
0 & 1 & 1\\
0 & 1  &0\\
\sqrt{2} &0& 1
\end{bmatrix}=
\begin{bmatrix}
0 & 1 & 1\\
0 & 1  &0\\
\sqrt{2} &0& 1
\end{bmatrix}
\begin{bmatrix}
0 & 1 &0\\
0 & 0  &0\\
0 &0& \frac{1}{\sqrt{2}}
\end{bmatrix}\nonumber.
\end{align}
Usually the matrix can be made normal by swapping the orbital order in either the bra or ket to maximize the diagonal terms of the orbital overlap.
That is, 
\begin{equation}
    \sum_{\nu\tau} \sum_k {^IC^\dagger_{j\nu}}g_{\nu\tau}\Big(\sum_i{^JC_{\tau i}}R_{ik}\Big)V_{kq}= \sum_i V_{ji}s_{iq},
\end{equation}
where $\mathbf{R}$ is a product of 90$^\circ$ rotations that maximizes the absolute trace of the matrix.
As the order is arbitrary, the only change is the reduced overlap sign which must be flipped if an odd number of swaps is performed.
Swapping the second and third columns of $\mathbf{A}$ results in a normal matrix and, therefore, a diagonal 
overlap matrix and a unitary eigenvector matrix,
\begin{align}
\begin{bmatrix}
\frac{1}{\sqrt{2}} & -\frac{1}{\sqrt{2}} &0\\
0 & 0  &0\\
\frac{1}{\sqrt{2}} &\frac{1}{\sqrt{2}}& 0
\end{bmatrix}
\begin{bmatrix}
1 & 0 & 0\\
0 & 0 & 1\\
0 & 1 & 0
\end{bmatrix}
\begin{bmatrix}
0 &\frac{-i}{\sqrt{2}} &  \frac{i}{\sqrt{2}}\\
1 & 0  & 0\\
0 & \frac{-i}{\sqrt{2}}  & \frac{1}{\sqrt{2}}
\end{bmatrix}=
\begin{bmatrix}
0 &\frac{-i}{\sqrt{2}} &  \frac{i}{\sqrt{2}}\\
1 & 0  & 0\\
0 & \frac{-i}{\sqrt{2}}  & \frac{1}{\sqrt{2}}
\end{bmatrix}
\begin{bmatrix}
0 & 0 &0\\
0 & \frac{(1-i)}{\sqrt{2}}  &0\\
0 &0& \frac{(1+i)}{\sqrt{2}}
\end{bmatrix}=
\begin{bmatrix}
0 &\frac{-i}{\sqrt{2}} &  \frac{i}{\sqrt{2}}\\
1 & 0  & 0\\
0 & \frac{-i}{\sqrt{2}}  & \frac{1}{\sqrt{2}}
\end{bmatrix}
\begin{bmatrix}
0 & 0 &0\\
0 & e^{-i\pi/4}  &0\\
0 &0& e^{i\pi/4}
\end{bmatrix}\label{apB:ex2}.
\end{align}
In both cases, the sign of reduced overlap is inverted, in SVD due to $\text{sgn}(\det(\mathbf{U}))$ and in the eigenvalue problem due to the odd number of rotations.
There are two drawbacks of this approach, first the transformation can be complex as shown in eq.~\eqref{apB:ex2}.
This increases the computational cost if methodology does not require complex orbitals, but there is no increase in cost if it does.
Second, the orbital order swapping is not guaranteed to produce a normal matrix.
In these cases, an arbitrary occupied-occupied rotation is necessary, which is similar to a SVD.
We point out that both approaches must lead to the same final result.

\section{Appendix C: implementation and scaling}

The calculation of matrix elements are implemented using the Generalized nonorthogonal Wick's theorem\cite{BurtonJCP21_154_144109,BurtonJCP22_157_204109} based on the natural excitation framework (NEF).\cite{moraesJCP26_164_014107}
Similar to other \TDMCSCF{} methods,\cite{satoPRA13_88_023402} the computational bottleneck at each time step is two-electron integrals projected over the single excited configurations required for the orbital time derivatives.
Using the Wick's theorem the scaling with respect to the number of reference configurations $N_\text{ref}$ and number of orbital basis $n$ is divided between the transformation of the two-electron integrals from atomic orbitals to the common basis, scaling as $\mathcal{O}(n^5)$, and the evaluation of the screened overlap matrices with $\mathcal{O}(N_\text{ref}^2n^3)$ scaling, and their partially contracted intermediate matrices with scaling $\mathcal{O}(N_\text{ref}n^4)$.\cite{BurtonJCP21_154_144109}
Therefore, for a fixed number of reference determinants the formal scaling is similar to MP2.
In this work, as the all common orbitals are considered active, the same integral transformation can be used for all time steps rather than a new transformation per iteration.
As a consequence, only the aforementioned screened overlap and contraction matrices must be recomputed at each time step, making the scaling with basis set size similar to HF.
If the common orbitals partition in occupied, active and virtual spaces is used instead, the number of excitations is reduced, but the integral transformation is required at every time step.

\bibliography{icmrnoci}

\end{document}